\documentclass[journal,10pt,twocolumn]{IEEEtran}
\usepackage[table]{xcolor}
\usepackage{balance}
\usepackage{multicol}
\usepackage{color, colortbl}
\usepackage[first=0,last=9]{lcg}

\usepackage{pdflscape}
\usepackage[english]{babel}
\usepackage{multicol}
\usepackage{multirow}
\usepackage{amsmath}
\usepackage{bm}
\usepackage{amsmath}
\usepackage{amsfonts}
\usepackage{amssymb}
\usepackage{float}
\usepackage{stfloats}
\usepackage{subfigure}
\usepackage{float}
\usepackage{amsmath}
\usepackage{color}
\usepackage{float}
\usepackage{array}
\usepackage{graphicx}
\usepackage{epstopdf}
\usepackage{epsfig}
\usepackage{algorithm}
\usepackage{textcomp}
\usepackage[noend]{algorithmic}
\usepackage{comment}
\usepackage[numbers, square, comma, sort&compress]{natbib}
\usepackage[numbers]{natbib}

\usepackage[singlespacing]{setspace}
\usepackage{etoolbox}
\usepackage{ccaption}
\usepackage{scrextend}
\usepackage{hyperref}
\usepackage[font=footnotesize,labelfont=rm,figurename=Fig.]{caption}
\makeatletter
\let\@nodottedtocline\@dottedtocline
\patchcmd{\@nodottedtocline}{\hbox{.}}{\hbox{}}{}{}
\patchcmd{\@nodottedtocline}{\normalcolor #5}{\normalcolor}{}{}
\newcommand*\l@sectionsubtitle{\@nodottedtocline{1}{0em}{1.5em}}
\makeatother

\newcounter{inlineequation}
\DeclareMathAlphabet{\mathpzc}{OT1}{pzc}{m}{it}
\definecolor{Gray}{gray}{0.9}
\definecolor{LightCyan}{rgb}{0.88,1,1}

\begin{document}
\title{From D2D to Ds2D: Prolonging the Battery Life of Mobile Devices via
Ds2D Communications}
\author{\IEEEauthorblockN{Muhammad Z. Shakir, Muhammad Ismail,  Xianbin Wang, Khalid A. Qaraqe, and Erchin Serpedin
\thanks{M. Z. Shakir is with the School of Engineering and Computing, University of the West of Scotland, Paisley, Scotland, UK, Email: muhammad.shakir@uws.ac.uk.}
\thanks{M. Ismail and K. A. Qaraqe are with Dept. of Electrical and Computer Engineering , Texas A\&M University at Qatar,  Doha, Email: \{m.ismail, khalid.qaraqe\}@qatar.tamu.edu.}
\thanks{X. Wang is with Dept. of Electrical and Computer Engineering, University of Western Ontario, Canada, Email: xianbin.wang@uwo.ca.}
\thanks{E. Serpedin is with Dept. of Electrical and Computer Engineering,  Texas A\&M University, College Station, TX, Email: serpedin@ece.tamu.edu.}
\thanks{Part of this work has been published in book ``Green Heterogeneous Wireless Networks," published jointly by Wiley and IEEE, Sep. 2016.}
}

}
\maketitle
\thispagestyle{empty}

\begin{abstract}

Emerging device centric systems (DCS) such as device-to-device (D2D)
communications are considered as a standard part of future mobile
networks, where operators/consumers involve the devices in direct
communication to improve the cellular system throughput, latency,
fairness, and energy efficiency. However, battery life of mobile
devices involved in such communications is crucial for 5G smartphone
users to explore the emerging applications in DCS. It is
anticipated that the owners of 5G-enabled smartphones use their
devices more extensively to  talk, text, email, and surf the Web
more often than do customers with 4G smartphones or traditional
handsets, which puts a significantly higher demand on the battery
life. Smartphones are currently equipped with multiple radio
interfaces that enable them to access different types of wireless
networks including LTE-direct and Wi-Fi-direct, besides cellular networks. Such a capability is not well explored within the context of DCS. This article proposes a new scheme to support the emerging features in DCS where a D2D-enabled mobile device (sink device or a file/content requester) aggregates the radio resources of multiple mobile devices (source devices or
file/content providers) via its multiple radio interfaces such that the scheme is referred to as devices-to-device (Ds2D) communications. Ds2D communication scheme ensures an optimal packet split among the source mobile devices to improve the file/content transfer latency (FTL), energy efficiency, and battery life. Simulation results demonstrate that the proposed optimal packet split scheme among multiple source devices participating in Ds2D communication scheme guarantees an improvement in mobile battery life over wide range of data rate levels in comparison with the random packet split strategy and traditional D2D communication paradigm between the sink and source
mobile devices.

\end{abstract}



\section{Introduction}

The recent widespread use of mobile Internet complemented by the
advent of many smart applications has led to an explosive growth in
mobile data traffic over the last few years. This remarkable growing
momentum of the mobile traffic will most likely continue on a similar
trajectory, mainly due to the emerging need for connecting people,
machines, and applications in an ubiquitous manner through the
mobile devices. Every new release of iPhone and Android smartphone
spurs new applications and services, with advanced display screens
to deliver an exceptional quality of experience to the end-user. As
a result, the current and projected dramatic growth of mobile data
traffic necessitates the development of the fifth-generation (5G) mobile
communications technology. The 5G communications are expected to yield a mobile broadband experience far beyond the current 4G systems. 

The 5G has a broad vision and envisages design targets that include 10-100x peak date rate, 1000x network capacity, 10x energy efficiency, and 10-30x lower latency \cite{ref82}. In achieving these expectations, operators and carriers are planning to leverage emerging device centric systems (DCS) such as
device-to-device (D2D) communications, small cells, nano and elastic
cells to improve the user experience and consequently improve the
overall network performance. However, the evolution of mobile
devices to support the emerging features in DCS comes at a cost placing stringent demands on the mobile device battery life and energy
consumption \cite{handset}. Hence, there are considerable market
interests on the development and deployment of  innovative green and
smart solutions to support emerging features in DCS  in ultra-dense
heterogeneous networks.

This article focuses on D2D communication systems and their architecture as an effective means for DCS to support the expectations of 5G networks.
The challenges  related to the implementation of D2D communications,
including interference management, energy consumption, and channel
measurements are briefly discussed. A new devices-to-device (Ds2D)
communication paradigm  is then proposed to establish D2D communication over multiple radio interfaces between a sink and source devices and thereby improve  energy consumption and battery life of mobile devices involved in such emerging
DCSs. The performance of the proposed emerging device centric
framework is then studied quantitatively, followed by an overview of
the challenges related to the implementation and integration
perspectives of Ds2D communications under 5G centralized and
decentralized network approaches. Finally, conclusions and 
future directions are outlined.


\section{Emerging Device Centric Paradigms}

So far, most of the deployed telecommunications networks (3G and 4G) have assumed a network centric approach. However, the new 5G systems are expected to drop this vision and adopt a DCS strategy \cite{5g, 7063545}. It is envisioned that the 5G networks would be mostly deployed for data-centric applications rather than voice-centric applications. The main drivers of DCS are Internet of Things (IoT), Machine-to-Machine communications and BigData applications, which
will exploit the intelligence at the mobile device side to support
the emerging device centric communication paradigms and ensure
ubiquitous connectivity.

\subsection{Overview of D2D Communication Architecture}

\begin{figure}
\begin{center}
\includegraphics[width=0.4\textwidth,bb=148 81 349 257]{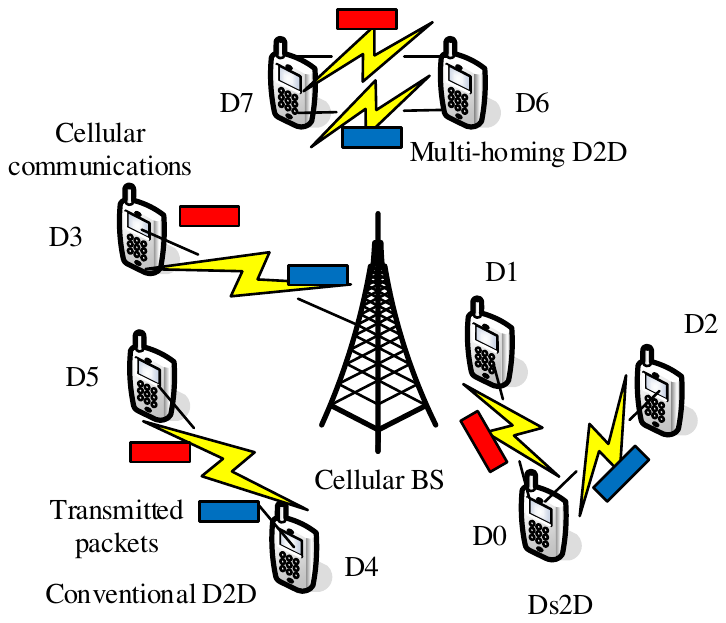}
\caption{Illustration of traditional D2D, multi-homing D2D, and Ds2D
communication approaches.} \label{system}
\end{center}
\end{figure}

D2D communication is viewed as a promising technology  to
complement the 5G DCS. Direct D2D communication between cellular
equipment is proposed to increase data-rate \cite{7590037}, extend conventional cellular coverage \cite{ref69,ref70} and wireless sharing/dissemination of content \cite{6829954}. As shown in Fig. \ref{system},  traditional D2D
communications take place among two devices, i.e., a pair of devices
($D_4$ and $D_5$) such that a direct communication link is
established between the two mobile devices without any interaction
from the base stations or the core of the cellular network. Hence,
the sink device ($D_5$) receives the required file directly from the
source device ($D_4$).

In~\cite{ref64}, the authors have provided a literature review on
D2D communications including new insights concerning existing works
and emerging protocols.  This study offers a review on the inband (underlay or overlay in cellular spectrum) and outband (unlicensed spectrum) integration of D2D communications. In an underlay scheme, the D2D communication may
generate interference to the cellular users due to the reuse of the
same resources. Hence, in the underlay approach, D2D links may only
exist if they do not harm the signal-to-interference plus
noise-ratio (SINR) at the base stations (uplink) or at the other
devices (downlink) in the conventional cellular communication
approach. 
Outband D2D communication uses cellular interface to set-up the connection between two devices for D2D communication and uses the Wi-Fi interface for data
transmission.

In~\cite{ref71}, the authors have proposed a new LTE-A-based D2D communication network architecture. They have introduced a new reference point between D2D-enabled devices named ``Di-interface" using enhanced radio protocols and procedures as a device connecting to eNodeB. The following D2D-specific functionalities are supported by many functions of this interface: first, the D2D scheme should have the ability to measure the distance between two mobile devices in order to relieve the feasibility of direct connection; second, the devices in the D2D architecture should be covered by the eNodeBs to maintain control and signaling; third, D2D data transmission between the devices should utilize a physical channel similar to the LTE-A uplink/downlink shared channel.




\subsection{D2D Communication Challenges}
Some challenges to implement and integrate device centric communications into 5G networks  are listed below~\cite{ref64}:

\begin{itemize}
\item Interference and Resource Management: For the reuse of uplink and downlink resources in D2D communications in a cellular network, the D2D mechanism should be designed in a way not to disrupt the cellular network services. The transmission power should be properly regulated so that the D2D transmitter does not interfere with the cellular mobile user communication while maintaining a minimum SINR requirement for the D2D receiver.
\item Channel Measurement/Modulation Format: D2D communication requires information on the channel gain between D2D pairs, the channel gain between D2D transmitter and cellular device, and the channel gain between cellular transmitter and D2D receiver. As the devices are supposed to communicate with both base stations and other peers, it is convenient to preserve some sort of common physical layer waveform such as orthogonal frequency-division multiplexing (OFDM) modulation.
\item Energy Consumption: While energy consumption is a very important issue in D2D communications, it becomes very crucial to propose advanced device discovery, device pairing, and D2D communication protocols to prolong the battery life of the mobile devices while ensuring the required QoS and connectivity.
\end{itemize}

\section{Devices-to-Device (Ds2D) Communications}

The opportunity of enabling multiple radio interfaces including LTE-direct and Wi-Fi-direct, besides cellular networks is not fully exploited in D2D communications since the D2D communication takes place over a single link between the two mobile devices involved in such direct communications (as shown in Fig. \ref{system} between $D_4$ and $D_5$). Communication over multiple links can take advantage of the diverse resources available at different radio interfaces (e.g., the supporting bandwidth). Aggregating such radio resources at the sink device allows for an improved system performance in terms of the achieved throughput, latency, and energy efficiency. Enabling D2D data transmissions over multiple radio interfaces can take place in two forms:
\begin{itemize}
\item Multi-homing D2D communications, in which the sink device receives its required file from a single source device over multiple radio interfaces, as shown in Fig. \ref{system} between the source device $D_6$ and the sink device $D_7$. For sake of illustration, assume that $D_7$ requests a file that consists of $2$ packets from $D_6$. Two data communication links are established between $D_6$ and $D_7$, which can take place over the LTE-direct and Wi-Fi-direct radio interfaces of the two devices (besides a third (cellular) link that is established for coordination). Based on the achieved data rate over each link (radio interface), different number of packets can be transmitted on each link from $D_6$ to $D_7$. For instance, one set of the data packets is transmitted over the first link and another set of the data packet is transmitted over the second link, as shown in Fig. \ref{system}. Eventually, the sink device $D_7$ aggregates the received two set of packets to reconstruct the required file.

\item Ds2D communications, in which the sink device receives its required file from multiple source devices over multiple radio interfaces, as shown in Fig. \ref{system} between the source devices $D_1$, $D_2$, and the sink device $D_0$. One data communication link is established between each source device and the sink device over different radio interfaces. For instance, data communication can take place between $D_1$ and $D_0$ over the LTE-direct radio interface and between $D_2$ and $D_0$ over the Wi-Fi-direct radio interface (besides a second (cellular) link that is established between each source device and the sink device for coordination). Again, based on the achieved data rate over each link (radio interface), different number of packets can be transmitted from each source device $D_1$ and $D_2$ to $D_0$. In Fig. \ref{system}, one set of the data packets is transmitted from $D_1$ and another set of the data packets is transmitted from $D_2$ and the sink device $D_0$ aggregates the received two set of the packets to reconstruct the required file.
\end{itemize}

One obvious advantage of multi-homing D2D over Ds2D communications
is the simple communication architecture since it involves
coordination between only two devices over multiple radio
interfaces. On the other hand, Ds2D communications involves
coordination among multiple devices over multiple radio interfaces.
However, from an energy efficiency perspective, Ds2D communications
offer a better alternative over multi-homing D2D, as presented in
our preliminary results in \cite{ds2d}. Specifically, in
multi-homing D2D, a single source device incurs energy consumption
for data communication over $N$ radio interfaces, while in Ds2D
communications, each of the $N$ source devices consume energy for
data communication over a single active radio interface. Furthermore, in
Ds2D communications, each source device transmits only a portion of
the required file, while in multi-homing D2D, the source device
transmits solely the entire file to the sink device. Hence, roughly,
Ds2D communications incur $1/N$ of the multi-homing D2D
power consumption per source device. In addition, Ds2D
communications incur lower transmission power per source device,
compared with the multi-homing D2D and traditional D2D approach,
based on the file split ratio among the different source devices.
Such an improved energy consumption per source device prolongs the
source device battery lifetime as compared with traditional and
multi-homing D2D and thus provides a better incentive for mobile
devices to participate in a direct communication transaction. In the
following, we focus on Ds2D communications among multiple source
devices over multiple radio interfaces to establish the links with a single sink device for data transmission and discuss the performance of the
communication framework along with the associated challenging issues.

\subsection{System Model and Network Layout}

Consider a system model with a single sink mobile device and a set
of candidate source mobile devices. Here, sink mobile device
requires to download a popular file (content) which is cached in
source mobile devices. Let $\bm{\mathcal{D}} = \{D_0, D_1, \ldots,
D_S\}$ denote a set of mobile devices with $D_0$ representing the
sink mobile device and $D_s \in \bm{\mathcal{D}} \setminus \{D_0\}$
representing the candidate source mobile devices. All mobile devices
in $\bm{\mathcal{D}}$ are in the coverage area of a single cellular
network base station. The candidate source mobile devices are
selected by the base station based on two criteria: 1) all devices
in $\bm{\mathcal{D}}$ are within the proximity of the sink mobile
device and 2) each candidate source mobile device has a copy of the
file required by the sink mobile device. Each mobile device $D_s \in
\bm{\mathcal{D}}$ has a set of distinct radio interfaces
$\bm{\mathcal{N}} = \{1, 2, \ldots, N\}$. Radio interface $n \in
\bm{\mathcal{N}}$ in all mobile devices $D_s \in \bm{\mathcal{D}}$
employs the same access technology. For instance, $n = 1$ represents
an LTE-direct radio interface, $n=2$ represents a Wi-Fi-direct radio
interface, $n = 3$ represents cellular radio interface in all
devices, etc. The Ds2D communication framework involves two phases,
namely, the optimal selection of source devices and optimal packet split
among source devices, which are discussed next.

\subsection{Optimal Selection of Source Devices}

In Ds2D communications, the base station is required to select the
optimal source mobile devices from the available candidate source
mobile devices and their respective radio interfaces that deliver
the required data (file) to the sink mobile device in the most
energy efficient manner.


The optimal algorithm for source mobile device and radio interface selection should maximize the achieved energy efficiency while accounting for some constraints. One constraint should ensure that the total number of links used for data transmission is limited by the maximum number of available radio interfaces at the sink device. Another constraint should guarantee that each radio interface of the sink device is communicating with only one source device. Furthermore, each source device should perform data transmission using a single radio interface. In \cite{ds2d},  an optimal algorithm for source mobile device and radio interface selection based on the ascending proxy auction mechanism was presented. The proposed mechanism achieves (i) higher energy efficiency compared with the traditional D2D communications approach and (ii) lower energy consumption per source mobile device compared with the multi-homing D2D communications approach.  To avoid triggering source device and radio interface selection every time a change occurs in the channel condition, and hence avoid high signaling overhead, the objective of the proposed algorithm is based on maximization of the time-average energy efficiency. 

\subsection{Optimal Packet Split among Devices}

After optimal selection of source mobile devices and their
respective radio interfaces, the base station coordinates with the
source mobile devices  to transfer the desired data packets to the
sink mobile device in a distributed manner. The sink mobile device
aggregates the data packets transmitted by different source mobile
devices to reconstruct the required file. This approach can support
data hungry applications such as file download or video streaming of
a popular content.

The optimal packet split algorithm should specify the packet split
ratio among the source devices based on the achieved data rates over
their respective radio interfaces. Consider that the desired file
(content) has $P$ long data packets that should be transmitted from
the source mobile devices (e.g., $D_1$ and $D_2$ as shown in Fig.
\ref{system}) to the sink mobile device ($D_0$) over a set of two
different radio interfaces $\bm{\mathcal{N}} =\{1,2\}$, as shown in
Fig. \ref{system}. Let $0 < \alpha_{\mbox{\footnotesize{opt}}}\leq
1$ denote the optimal packet split ratio (OPSR) that splits the
requested file into two sets of data packets based on the achieved
data rate for each selected source device. Set 1 of data packets
contains $\alpha_{\mbox{\footnotesize{opt}}}\:P$ data packets that
are transmitted by source mobile device $D_1$ through radio
interface $n =1$. Similarly, set 2 of data packets contains
$(1-\alpha_{\mbox{\footnotesize{opt}}})\:P$ remaining data packets
that are transmitted by the source mobile device $D_2$ through radio
interface $n =2$.  The sink mobile device receives the packets from
both source mobile devices simultaneously over two different radio
interfaces ($n=1$ and $n=2$) and combines them to restore the
requested file (content).

The two source mobile devices $D_1$ and $D_2$ can transmit with
different data rates $R_1$ and $R_2$, respectively, depending on the
SINR of each source mobile device on the corresponding radio
interface\footnote{In this article, the relationship between the
SINR and the achieved data rates is adopted from \cite[Table
II]{sinr}. As an example, vector of data rates measured in kbps and
achieved over the range of SINR, is assumed as   \{213.3, 328.2,
527.8, 842.2, 1227.8, 1646.1, 2067.2, 2679.7, 3368.8, 3822.7,
4651.2, 5463.2, 6332.8, 7161.3 ,7776.6\} kbps.}. The file transfer latency (FTL) $t$ at the sink mobile device is defined as the duration
required to transfer the desired data packets from all source mobile
devices to the sink mobile device by aggregating the multiple radio
resources and is given by
\begin{equation}
t=\underset{n\in \bm{\mathcal{N}}}{\mbox{{max}}} \left\{ \frac{P_n\:B}{R_n}\right\}\:\:\: \text{[seconds]},
\label{1}
\end{equation}
where $P_n$ is number of data packets transmitted over the
$n^{\mbox{\footnotesize{th}}}$ radio interface (using $\alpha$, $P_1=\alpha\:P$ and $P_2=(1-\alpha)\:P$); $R_n$ denotes data
rate over the $n^{\mbox{\footnotesize{th}}}$ radio interface, and
$B=1500 \times 8$ denotes the bits per data packet by assuming that each data
packet contains 1500 bytes and each byte contains 8 bits. From
(\ref{1}), the FTL is minimum if all source
devices complete their data transmissions at the same time. Hence,
the main rationale behind the search of
$\alpha_{\mbox{\footnotesize{opt}}}$ is to ensure that the source
devices involved in Ds2D communications complete the file transfer
at the same time such that the sink mobile device does not have to
wait for one source mobile device to complete the  transmission of
its assigned data packets, which elongates the communication
sessions and leads to higher energy consumption and lower battery
life. Thus, $\alpha_{\mbox{\footnotesize{opt}}}$ can be found by
solving $(\alpha_{\mbox{\footnotesize{opt}}}\:P\:B)/(R_1) =
((1-\alpha_{\mbox{\footnotesize{opt}}})\:P\:B)/(R_2)$, i.e., $\alpha_{\mbox{\footnotesize{opt}}} = R_1/(R_1+R_2)$.


\begin{figure}
\begin{center}
\includegraphics[scale=0.55]{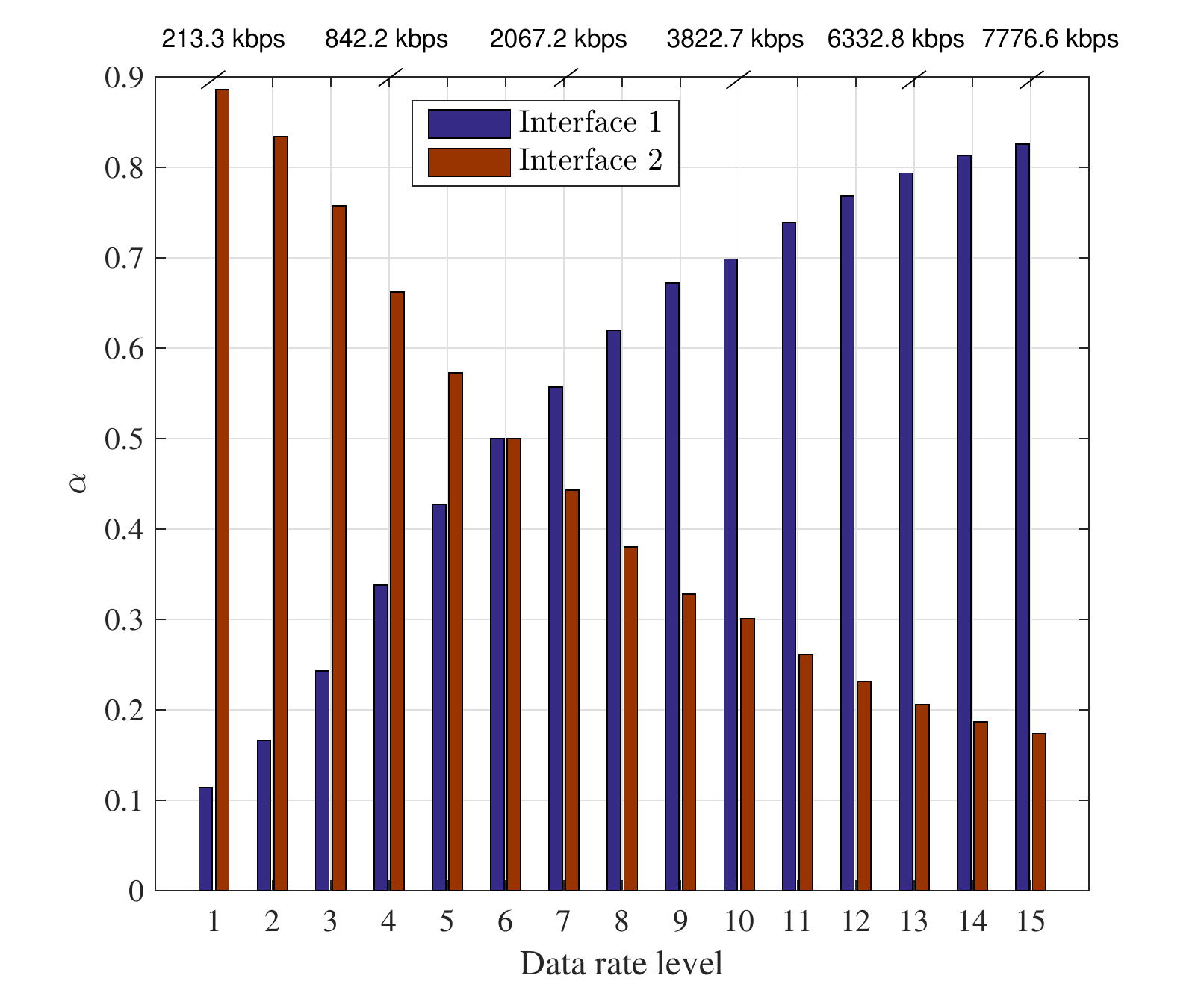}
\caption{Optimal packet split over two interfaces of two source mobile devices vs. range of data rate levels.}
\label{packet_split}
\end{center}
\end{figure}

\begin{figure}
\begin{center}
\includegraphics[scale=0.55]{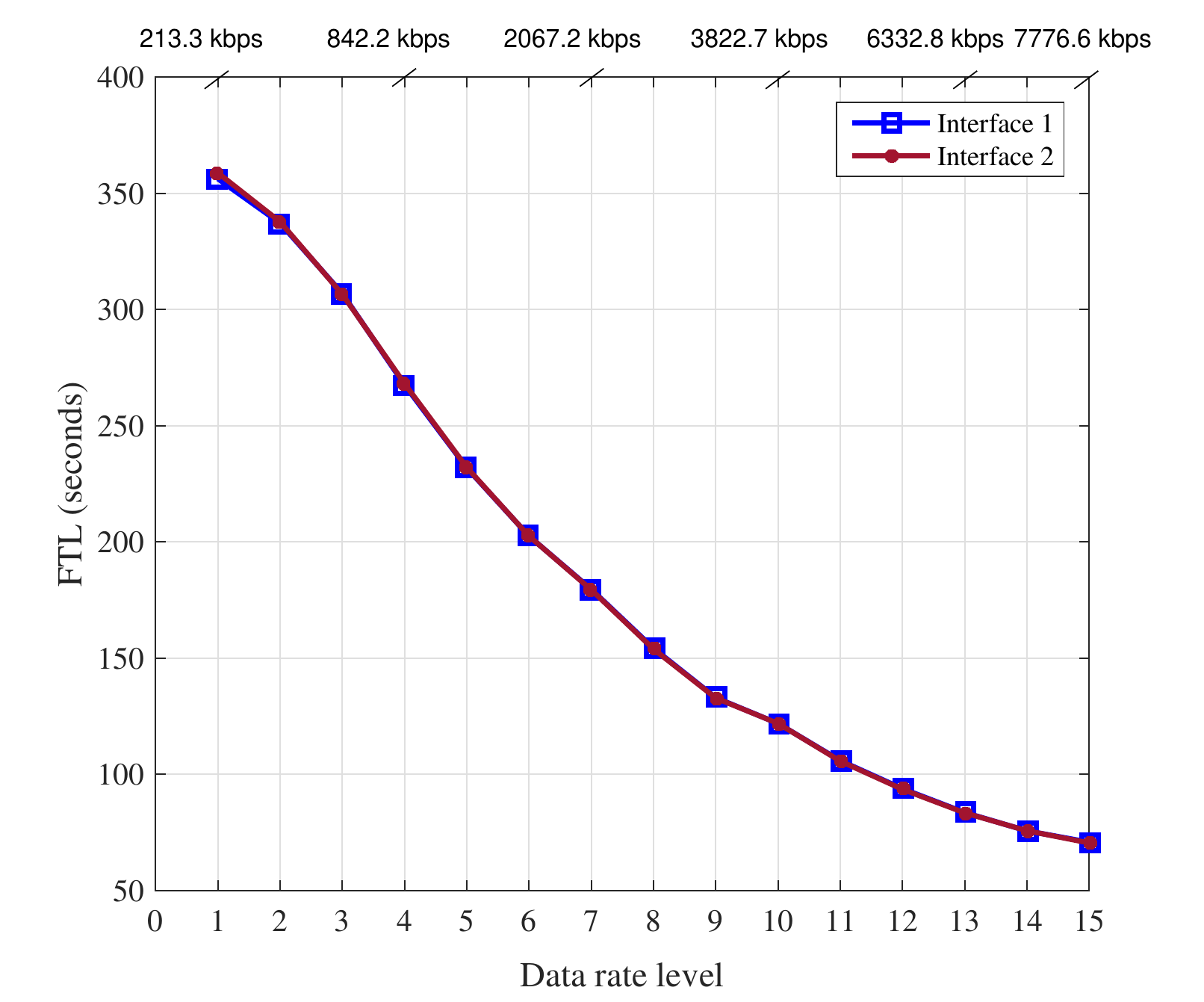}
\caption{Latency of transferring the requested file to sink mobile device over two radio interfaces of two source mobile devices by exploiting optimal packet split.}
\label{latency}
\end{center}
\end{figure}

To evaluate the effectiveness of the proposed optimal packet split
strategy over the two radio interfaces of two source mobile devices,
we consider an average monthly data usage capability for each mobile
subscriber of about $2.5$ GB with the daily download capability of
$80$ MB. Given each data packet has $1500$ bytes, the file
(requested content) has $P=55$k data packets. The average data rate
achieved for the second source mobile device ($D_2$) over radio
interface $n =2$ is assumed to be $1.646$ Mbps. Moreover, the
average achieved data rate for the first source device ($D_1$) over
radio interface ($n =1$) is varied for performance evaluation. Fig.
\ref{packet_split} shows the optimal packet split between a pair of
source devices ($D_1, D_2$) over two radio interfaces
$\bm{\mathcal{N}} =\{1,2\}$ for various achieved data rates of $D_1$.
The optimal packet split algorithm splits the data packets among the
two source mobile devices based on the achieved data rate for each
source mobile device. This is mainly because the optimal packet
split algorithm ensures the same FTL for each source mobile
device to guarantee the simultaneous delivery of the content/data packets to the sink device for content/data packets aggregation (as shown in the Fig. \ref{latency}).


\begin{figure}
\begin{center}
\includegraphics[scale=0.55]{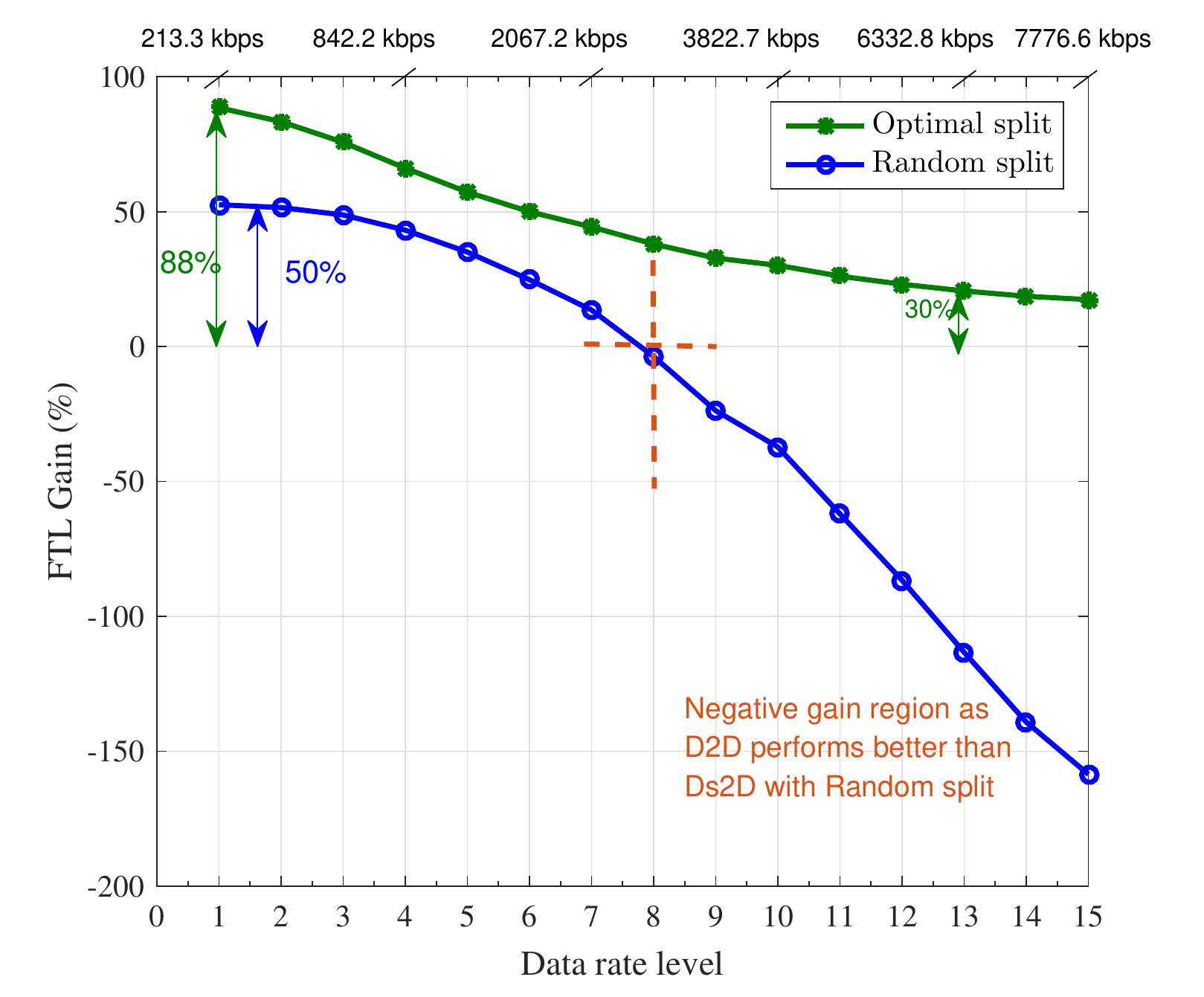}
\caption{Relative gain in file transfer latency (FTL) over Ds2D
communication with optimal packet split and random packet split in
comparison with traditional D2D communication.} \label{gain}
\end{center}
\end{figure}

Fig. \ref{gain} shows the relative percentage reduction in FTL (i.e., relative gain) for the considered Ds2D communication paradigms in comparison with the transmission over the traditional D2D communication paradigm where
only one source mobile device transmits the complete file to the
sink mobile device, i.e., direct D2D communication between a pair of
devices ($D_1$ and $D_0$). The performance of the Ds2D communication with optimal packet split algorithm is evaluated against Ds2D communication with random packet split benchmark.

The optimal packet split algorithm splits the file between the source file in such a way that guarantees the minimum FTL, i.e., data packets are distributed based on the channel conditions between the source devices and a sink device. On the contrary, The random packet split benchmark algorithm randomly splits the file among the two source mobile devices and each source mobile device transfers the packets to the sink mobile device that combines both sets to restore the requested file. It can be seen clearly that Ds2D transmission with optimal packet split has lower transmission FTL than the traditional D2D paradigm (there is always a gain, which ranges from 88$\%$ to 30$\%$). As shown in figure, with the increase in the data rate level for $D_1$, the achieved relative gain in FTL is reduced. This is mainly because with high data rates achieved for $D_1$, a single transmission link (between $D_1$ and $D_0$) can already achieve low FTL as compared with the Ds2D communication with optimal packet split algorithm (among $D_1$, $D_2$, and $D_0$). On the other side, in Ds2D transmission with random packet split, the relative gain in FTL can go below zero when a source device with worse channel condition is allocated more data packets for transmission compared with another source device with better channel conditions. In this case, the overall latency for Ds2D transmission with random packet split algorithm is increased compared with the single interface scheme (direct D2D transmission), which in turn deteriorates the relative gain of Ds2D with random packet split compared with the conventional D2D scheme and hence results in a gain below zero, i.e., a loss.


\begin{figure}
\begin{center}
\includegraphics[scale=0.55]{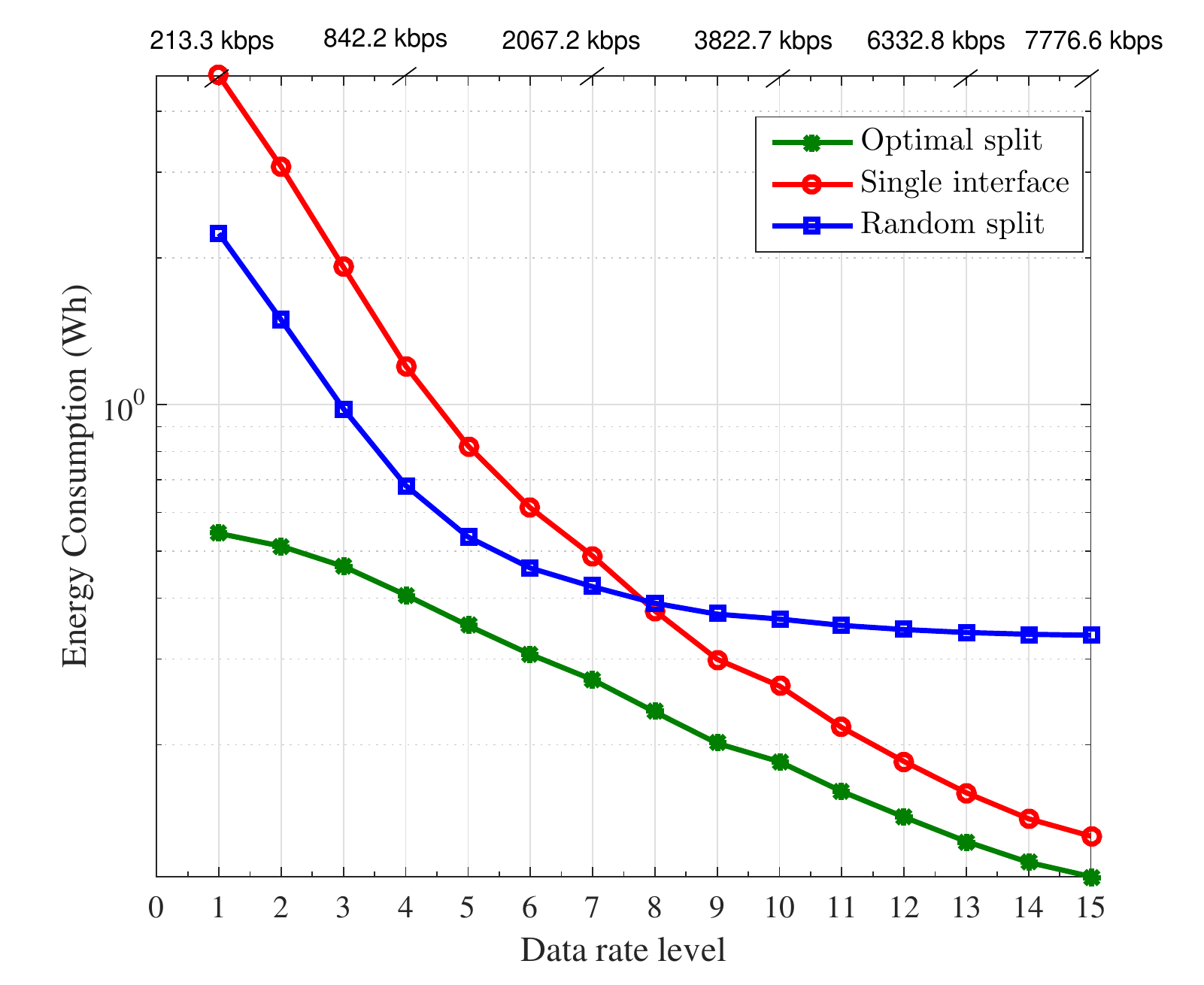}
\caption{Energy consumption (Wh)  per source device  involved in Ds2D communication with optimal and random packet split algorithms and D2D communication vs. range of achieved data rate. Energy consumption is only accounted for the active radio interface per device involved in Ds2D communication.} \label{power_consumption}
\end{center}
\end{figure}

\section{Green Analysis of Mobile Devices in Ds2D Communications}

In this section, we demonstrate the impact of the considered Ds2D communication on the society and consumers (mobile users) in terms of reduction in energy consumption, associated electricity cost and carbon footprint and improvement in mobile battery life. 

Let us consider a source mobile device such that its battery
holds a charge of $I_{\mbox{\footnotesize{batt}}}=1440$ mAh with
$E_{\mbox{\footnotesize{batt}}}=5.45$ Wh \cite{handset}. Energy
consumption of a source mobile device for transferring a file to a sink mobile device can be calculated as
\begin{equation}
E_{\mbox{\footnotesize{source}}}=\frac{E_{\mbox{\footnotesize{batt}}}\: t}{3600}\:\:\:\text{[Wh]},
\end{equation}
where $t$ is the FTL per source mobile device measured in seconds to transfer the assigned content to a sink mobile device, i.e., optimally and randomly assigned content in Ds2D and full content in D2D communications. 

Fig. \ref{power_consumption} shows the energy consumption per source
device  involved in transferring optimally assigned data packets out of the file of size $80$ MB (or equivalently $P=55$k data packets) to a sink device over the range of date rate levels. Compared with the direct D2D communication between a pair of devices, the proposed Ds2D communication offers reduced energy consumption per source mobile device since each source device only transmits a fraction of packets of the requested file. However, the energy consumption of the source mobile devices involved in Ds2D
communication under the optimal packet split scheme outperforms the
energy consumption of source mobile devices under the random packet
split scheme. Moreover, at a lower date rate level, the energy
consumption of the source mobile devices involved in Ds2D under the optimal
packet split scheme is significantly reduced in comparison with the
energy consumption of the source mobile devices involved in Ds2D
communication under the random packet split scheme and traditional  D2D
communications. The improvement is due to the fact that the source
mobile device is engaged with the sink mobile device for relatively
longer duration at a lower data rate level to complete the transfer
of required file under traditional D2D communications in comparison
with the source mobile devices involved in Ds2D communication. As an
example, at a rate  level 2, i.e., $R_2=328.2$ kbps, the source
devices can achieve $85\%$ reduction in energy consumption under the
optimal packet split scheme and $51\%$ reduction under the random packet
split scheme in comparison with the source device involved in D2D
communications. As shown in Fig. \ref{power_consumption}, Ds2D
communication with optimally assigned data packets exhibits a
closer performance to D2D communication for higher data rate levels.
This is due to the fact that with high data rates achieved by the
source mobile device, a D2D transmission link (e.g., between
$D_1$ and $D_0$) can also achieve a relatively low FTL.


\begin{figure*}
\begin{center}
\includegraphics[scale=0.6]{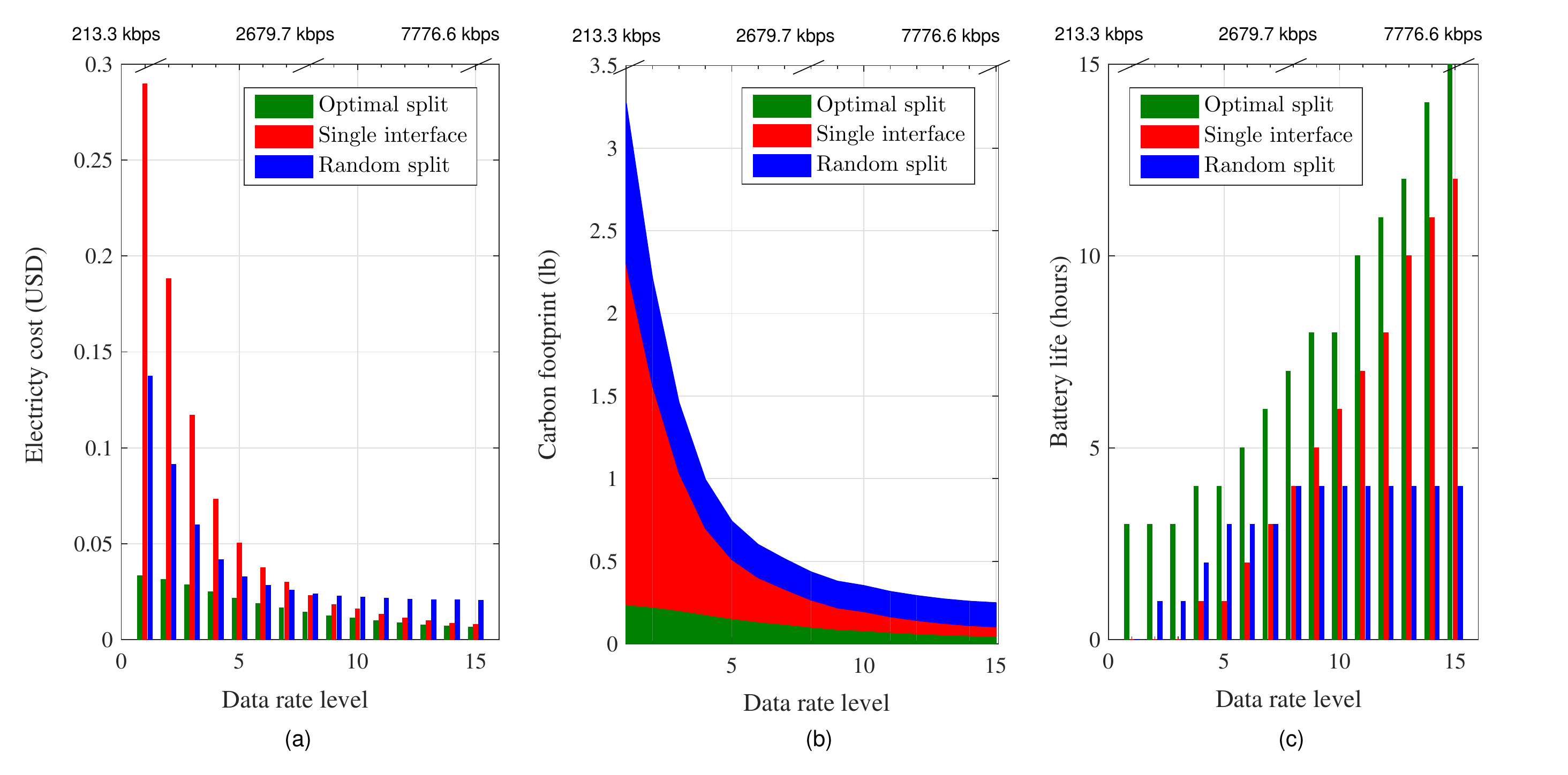}
\caption{Comparative summary of ecology and economic perspectives of various device centric frameworks for a mobile device  over the range of considered data rate levels: (a) annual electricity cost (USD); (b) carbon footprint (lb) and (c) battery life (hours). 
}
\label{cost}
\end{center}
\end{figure*}

Reducing energy consumption reduces the electricity cost and thereby
results in a financial cost saving to the consumers if the energy
savings offset any additional costs for implementing an energy
efficient framework. The monthly cost of electricity that is
associated with the implementation of considered device centric frameworks can be
calculated by assuming the cost of 1 kWh to be 12 cents: 
\begin{equation}
\text{Cost}=\frac{E_{\mbox{\footnotesize{source}}}}{1000 \times 100} \times 12 \:\:\: \text{[USD/month]},
\end{equation}

Fig. \ref{cost}a\footnote{Here, the electricity cost represents only
kWh spent for consuming the 2.5 GB monthly data allowance via
considered device centric frameworks without considering  other
associated additional electricity billing charges such as energy transmission and distribution charges.} shows the annual electricity cost associated with the energy  consumption of a source mobile device for the transfer of a $80$ MB file over the considered device centric frameworks. It can be seen clearly that
at an average price of $12$ cents per kWh, the mobile device costs
approximately $0.29$ USD in addition to the monthly electricity bill assumed by customers when a pair of devices is involved in D2D
communications with an average daily data usage or file/content
transfer of size $80$ MB at a lower data rate. On the contrary, the
electricity cost to the consumers reduced to $0.14$ USD and $0.04$
USD, when the devices are involved in Ds2D communication via the random
packet split and optimal packet split schemes, respectively, for
same amount of daily data transfer at a lower data rate. Again, with the
high data rates achieved by the source mobile device, a D2D
transmission link (e.g., between $D_1$ and $D_0$) can also offer a reduced
electricity cost as compared with the Ds2D communication with optimal packet split, which explains the close performance of Ds2D and D2D communications at high achieved data rates.


Energy efficiency has been recently marked as one of the alarming
bottlenecks in the telecommunication growth paradigm mainly due to
two major reasons, namely, (i) dramatically varying global
climate \cite{green} and (ii) slowly progressing  battery
technology \cite{battery}. In order to determine the ecological impact of the energy consumption of a mobile device due to the proposed framework, we calculate the corresponding carbon footprint  in pounds [lb]. The conversion factor used to convert the energy consumption to $\mathrm{CO_2}$ emissions is $1$ kWh $=1.21$ lb $\mathrm{CO_2}$ emissions which represents  the energy used at the mobile user's end \footnote{Energy and carbon conversions:fact sheet, available at \url{http://www.carbontrust.com}}. Fig. \ref{cost}b illustrates the annual $\mathrm{CO_2}$ emissions for a source device involved in 2.5GB monthly file transfer/download via (i)
D2D communications, (ii) D2s2D communications with random packet
split, and (iii) Ds2D communications with optimal packet split. It
can be seen clearly that the carbon footprint of mobile devices
involved in Ds2D communications is lower in comparison with the
carbon footprint of the devices involved in direct D2D communication.
Moreover, the carbon footprint of the devices is reduced significantly
when devices are communicating at relatively higher data rates. As
an example, at data rates of 213.3 kbps and 3822.7 kbps (data rate
level 1 and level 10), an annual $\mathrm{CO_2}$ emission of a source
mobile device involved in D2D communications is approximated as
being $2.06$ lb and $0.14$ lb, respectively.  D2sD communication
with random packet split between the devices reduces the estimated
$\mathrm{CO_2}$ emission of a involved source mobile device to $1.06 $ lb ($50 \%$ reduction) and $0.11$ lb (21 \% reduction) at the two considered data
rate levels, respectively. This can be further reduced  to $0.23$ lb
(more than $88\%$ reduction) and $0.08$ lb (42\% reduction),
respectively, by introducing an optimal packet split strategy between
the source mobile devices involved in Ds2D communication\footnote{The green analysis presented in this article is based on the carbon footprint of a single source mobile device involved in the device centric frameworks. However, with the advent of 5G communication paradigms and associated applications, it is anticipated that the carbon footprint of more than 6 billion mobile devices with an average 2.5 GB monthly data consumption via traditional D2D communication could reach 1 Mtonnes by 2020. This could be
reduced to 0.5 Mtonnes and 0.1 Mtonnes when the devices are
exploiting the Ds2D communications framework with random packet split and optimal packet split, respectively.}.


J. D. Power Associates demonstrates that the iPhone ranked top in all categories except for the battery life\footnote{Wireless smartphone customer satisfaction survey report, vol. 1, available at \url{http://www.jdpower.com/sites/default/files/2012030-whst.pdf}}. According to another recent survey report, up to 60\% of the mobile users in China complained that the battery consumption is the greatest hurdle while using 4G services\footnote{2010 China's Nokia mobile phone user research report, available at \url{http://zdc.zol.com.cn/201/2019387.html}}. Without a breakthrough in battery technology, the battery life of the mobile devices will remain the biggest limitation for energy-hungry device centric applications and services (e.g., video games, mobile P2P, interactive video, streaming multimedia, mobile TV, 3D services, and video sharing) \cite{cimini}. Emerging device centric frameworks can offer longer battery life while consumers can enjoy high data rate 5G services and applications. The battery life or battery capacity can be calculated from the input current rating of the battery and the load current of the battery charging circuit \cite{handset}. Battery life will be high when the load current is low and vice versa. The calculation to find out the capacity of battery can be mathematically expressed as\footnote{\label{first} Battery life calculator, available at \url{http://www.digikey.com/en/resources/conversion-calculators/conversion-calculator-battery-life}}

\begin{equation}
\text{Battery Life} = \frac{I_{\mbox{\footnotesize{batt}}}}{I_{\mbox{\footnotesize{source}}}} \times 0.70 \:\:\: \text{[hours]},
\end{equation}
where $I_{\mbox{\footnotesize{batt}}}$ is the battery capacity in
mAh, $I_{\mbox{\footnotesize{source}}}$ is the load current drawn by
the source mobile device for transferring the file to the sink
mobile device. Here, the factor $0.70$ represents external factors
that can affect the mobile device battery life\footref{first}. Fig.
\ref{cost}c shows the mobile battery life (hours) over the range of data
rate levels for a source mobile device involved in D2D and Ds2D communications. Overall, as the data rate level increases, the FTL is decreased, and hence the battery life is prolonged. However, it can be seen clearly that the battery life of a source mobile device
involved in Ds2D communications with an optimal packet split scheme
is significantly higher than  the battery life of a mobile device
involved in Ds2D communication with the random packet split scheme and
traditional D2D communication. Moreover, the battery life of the source
mobile device involved in Ds2D communication with random packet
split scheme is degraded at the higher data rate levels since the
FTL performance of the random packet split scheme is worse than the
traditional D2D communications (as can be seen from Fig.
\ref{latency}). As an example, at a low data rate of 213.3 kbps (data
rate level 1), the battery life of a mobile device involved in Ds2D
communication with optimal packet split prolongs to approximately 3
hours. On the contrary, a fully charged battery is not enough to
successfully complete the file transfer under Ds2D communication
with the random packet split scheme and traditional D2D communications.

\section{Some Challenges and Future Directions}


In general, Ds2D communications can be established among any sink mobile device and multiple source mobile devices over $N$ multiple radio interfaces. Selection of source mobile devices is highly dependent on the availability of the file or content and its close proximity with the sink mobile device. As discussed earlier, coordination among the involved  mobile devices is required for successful implementation of Ds2D communication and optimal distribution of the desired content (data packets) among the source mobile devices. There are two possible implementation approaches to achieve the coordination among the mobile devices and set-up Ds2D communication, namely, centralized and decentralized approaches, and are described as below\footnote{Centralized and decentralized approaches are first introduced in 3GPP Release 12 to ensure efficient radio resource allocation for D2D communication and are later integrated with LTE in 3GPP Release 13. More information can be seen in 3GPP technical report, Evolution of LTE in 3GPP Release 13, Feb, 2015, available at \url{ http://www.3gpp.org/news-events/3gpp-news/1628-rel13}}.

\subsection{Centralized Ds2D Set-up:} Under the centralized Ds2D set-up, cloud radio access networks (CRAN) can dynamically perform source mobile device selection, Ds2D link establishment, and data packet distribution among the source device with a limited or full supervision of cellular network. Devices involved in Ds2D communication can perform full or limited information exchange and signaling with the cellular network using the LTE-Uu interface (i.e., cellular link). Since cellular interface for all devices is reserved for information exchange and signaling, data transmission can be established between a sink mobile device and source devices over $N - 1$ radio interfaces. Therefore, the devices have at least two active interfaces (cellular interface for control and an additional radio interface for data transmission). Mobility of the devices involved in Ds2D communication, interference management, and content availability are considered as advantages of integrating Ds2D communications under centralized CRAN-enabled cellular systems. Inter-network coding can play an important role to efficiently exploit the benefits of Ds2D communications. However, the centralized approach imposes additional challenges to the fronthaul requirements, such as high data rate and latency requirements, due to  information exchange and signaling overheads between the devices involved in Ds2D communications. Moreover, devices cannot establish Ds2D communication links without full or limited intervention and approval to the request from the cellular network.

\subsection{Decentralized Ds2D Set-up:} Under the decentralized Ds2D set-up, devices involved in Ds2D communications can autonomously exchange control signaling for selection of source mobile devices, Ds2D communication establishment, and content distribution among the devices without any intervention from the cellular networks. Therefore, the devices can establish Ds2D communication over a relatively short time period under the decentralized system in comparison with the time required to set-up Ds2D links in a centralized manner. The cellular network does not have any supervision over the functionalities used by the devices involved in Ds2D communication, such as resource allocation, interference management, etc. Devices can use the PC5 interface, which is defined by the LTE standard for device discovery and D2D communication between users.  Moreover, fronthaul requirements can be relaxed due to a reduced signaling information exchange between the devices and access network. Long-term availability of the desired content due to the mobility is one of the challenges to integrate Ds2D communications in decentralized fashion.

\section{Conclusions}

Smartphones will play an important role to enable device centric communication paradigms in 5G networks, such as D2D communications. This article has focused  on the  implementation perspectives of such a device centric architecture including energy consumption and battery life of the devices involved in communication. A new device centric scheme, Ds2D communication, has been proposed in which a sink device aggregates multiple resources to download the desired content (file) from multiple source mobile devices. Ds2D communication guarantees an optimal data packet distribution among the source mobile devices to ensure improvements in file transfer latency, energy consumption, and battery life of the source mobile devices involved  in communications. Simulation results have evaluated the quantitative gains in comparison with the traditional D2D communication and Ds2D communication in which a random data packet distribution has been implemented. It has been shown that the performance criteria of the source mobile devices such as file transfer latency, energy consumption, and battery life can be effectively optimized through an optimal packet split among the source mobile devices and their respective radio interfaces involved in the Ds2D communication in DCS.

\balance

{\bibliography{IEEEfull,references}
\bibliographystyle{IEEEtran}}
\begin{IEEEbiographynophoto}{Muhammad Zeeshan Shakir} (S'04, M'10, SM'16) is an Assistant Professor (Lecturer UK) in Networks in the School of Engineering and Computing at University of the West Scotland (UWS), UK where he is a member of research institute for Artificial Intelligence, Visual Communication and Networks (AVCN). Before joining UWS in Fall 2016, he has been working at Carleton University, Canada, Texas A\&M University, Qatar and KAUST, Saudi Arabia on various national and international collaborative projects with Huawei, TELUS, NSERC, and QNRF. 

Dr. Shakir's main research interests lie in design and development of heterogeneous networks for urban and rural coverage via unconventional aerial and regular architectures.  He is also interested in developing unified integrated user centric frameworks for emerging 5G technologies such as Caching, D2D, MTC, IoT and SDR. 
Dr. Shakir has a track record of more than 75 technical journal and conference publications including many in the world's most reputable journals in the areas of wireless communications.

Dr. Shakir has been/is giving tutorials on emerging wireless communication systems at IEEE flagship conferences. 
He has been/is also serving as a Chair/Co-Chair of several workshops/special sessions.
He has been also serving on the technical program committee of different IEEE conferences, including Globecom, ICC, and WCNC. 
He is an Associate Technical Editor of IEEE Communications Magazine and has served as a Guest Editor for IEEE Wireless Communications and IEEE Access. Dr. Shakir is serving as a Chair of IEEE ComSoc emerging technical committee on backhaul/fronthaul communications and networking. He has been serving as an active member to several IEEE ComSoc technical committees. 
He is a Senior Member of IEEE, an active member of IEEE ComSoc and IEEE Standard Association.
\end{IEEEbiographynophoto}

\begin{IEEEbiographynophoto}{Muhammad Ismail} received the B.Sc. and M.Sc. degrees (Hons.) in electrical engineering (Electronics and Communications) from Ain Shams University, Cairo, Egypt, in 2007 and 2009, respectively, and the Ph.D. degree in electrical and computer engineering from the University of Waterloo, Waterloo, ON, Canada, in 2013. He is currently an Assistant Research Scientist with the Electrical and Computer Engineering Department, Texas A\&M University at Qatar, Doha, Qatar and an IEEE Senior Member. Dr. Ismail is a Co-Author of two research monographs by Wiley-IEEE Press and Springer. He is a co-recipient of the best paper awards in the IEEE ICC 2014, the IEEE Globecom 2014, the SGRE 2015, and the Green 2016. Dr. Ismail was an Editorial Assistant of the IEEE TRANSACTIONS ON VEHICULAR TECHNOLOGY from 2011 to 2013. He has been an Associate Editor of the IEEE TRANSACTIONS ON GREEN COMMUNICATIONS AND NETWORKING since 2016 and the IET Communications since 2014.
\end{IEEEbiographynophoto}

\begin{IEEEbiographynophoto}{Xianbin Wang} (S'98, M'99, SM'06, F'17) is a Professor and Canada Research Chair at Western University, Canada. He received his Ph.D. degree in electrical and computer engineering from National University of Singapore in 2001. Prior to joining Western, he was with Communications Research Centre Canada (CRC) as a Research Scientist/Senior Research Scientist between July 2002 and Dec. 2007. From Jan. 2001 to July 2002, he was a system designer at STMicroelectronics, where he was responsible for the system design of DSL and Gigabit Ethernet chipsets.  His current research interests include 5G technologies, signal processing for communications, adaptive wireless systems, communications security, and locationing technologies. Dr. Wang has over 280 peer-reviewed journal and conference papers, in addition to 26 granted and pending patents and several standard contributions.

Dr. Wang is a Fellow of IEEE and an IEEE Distinguished Lecturer. He has received many awards and recognition, including Canada Research Chair, CRC President's Excellence Award, Canadian Federal Government Public Service Award, Ontario Early Researcher Award and five IEEE Best Paper Awards. He currently serves as an Editor/Associate Editor for IEEE Transactions on Communications, IEEE Transactions on Broadcasting, and IEEE Transactions on Vehicular Technology and He was also an Associate Editor for IEEE Transactions on Wireless Communications between 2007 and 2011, and IEEE Wireless Communications Letters between 2011 and 2016. Dr. Wang was involved in a number of IEEE conferences including GLOBECOM, ICC, VTC, PIMRC, WCNC and CWIT, in different roles such as symposium chair, tutorial instructor, track chair, session chair and TPC co-chair.
\end{IEEEbiographynophoto}

\begin{IEEEbiographynophoto}{Khalid A. Qaraqe} (S'00) was born in Bethlehem. Dr Qaraqe received the B.S. degree in EE from the University of Technology, Bagdad, Iraq in 1986, with honors. He received the M.S. degree in EE from the University of Jordan, Jordan, Amman, Jordan, in 1989, and he earned his Ph.D. degree in EE from Texas A\&M University, College Station, TX, in 1997. From 1989 to 2004 Dr Qaraqe has held a variety positions in many companies and he has over 12 years of experience in the telecommunication industry. Dr Qaraqe has worked on numerous GSM, CDMA, and WCDMA projects and has experience in product development, design, deployments, testing and integration.  Dr Qaraqe joined the department of Electrical and Computer Engineering of Texas A\&M University at Qatar, in July 2004, where he is now a professor. Dr Qaraqe research interests include communication theory and its application to design and performance, analysis of cellular systems and indoor communication systems. Particular interests are in mobile networks, broadband wireless access, cooperative networks, cognitive radio, diversity techniques and 5G systems.
\end{IEEEbiographynophoto}

\begin{IEEEbiographynophoto}{Erchin Serpedin} (F'13) received the Specialization degree in signal processing and transmission of information from Ecole Superieure D'Electricite, Paris, France, in 1992, the M.Sc. degree from the Georgia Institute of Technology, Atlanta, GA, USA, in 1992, and the Ph.D. degree in electrical engineering from the University of Virginia, Charlottesville, VA, USA, in 1999. He is currently a Professor with the Department of Electrical and Computer Engineering, Texas A\&M University, College Station, TX, USA. He is the author of two research monographs, one textbook, nine book chapters, 110 journal papers, and 180 conference papers. His research interests include signal processing, biomedical engineering, bioinformatics, and machine learning. He was a recipient of numerous awards and research grants. He served as an associate editor of a dozen of journals, such as the IEEE TRANSACTIONS ON INFORMATION THEORY and the IEEE TRANSACTIONS ON SIGNAL PROCESSING, and as a technical chair for five major conferences. He is currently serving as an Associate Editor of the IEEE Signal Processing Magazine and as the Editor-in-Chief of the EURASIP Journal on Bioinformatics and Systems Biology, an online journal edited by Springer.
\end{IEEEbiographynophoto}

\end{document}